# Embedding Representation of Academic Heterogeneous Information Networks Based on Federated Learning

**Junfu Wang[1], Yawen Li[2*], Meiyu Liang[1], Ang Li[1]**

[1] Beijing Key Laboratory of Intelligent Communication Software and Multimedia, School of Computer Science (National Pilot Software Engineering School), Beijing University of Posts and Telecommunications, Beijing 100876
[2] School of Economics and Management, Beijing University of Posts and Telecommunications, Beijing

**Abstract:** Academic networks in the real world can usually be portrayed as heterogeneous information networks (HINs) with multi-type, universally connected nodes and multi-relationships. Some existing studies for the representation learning of homogeneous information networks cannot be applicable to heterogeneous information networks because of the lack of ability to issue heterogeneity. At the same time, data has become a factor of production, playing an increasingly important role. Due to the closeness and blocking of businesses among different enterprises, there is a serious phenomenon of data islands. To solve the above challenges, aiming at the data information of scientific research teams closely related to science and technology, we proposed an academic heterogeneous information network embedding representation learning method based on federated learning (FedAHE), which utilizes node attention and meta path attention mechanism to learn low-dimensional, dense and real-valued vector representations while preserving the rich topological information and meta-path-based semantic information of nodes in network. Moreover, we combined federated learning with the representation learning of HINs composed of scientific research teams and put forward a federal training mechanism based on dynamic weighted aggregation of parameters (FedDWA) to optimize the node embeddings of HINs. Through sufficient experiments, the efficiency, accuracy and feasibility of our proposed framework are demonstrated.

## 1 Introduction

With continuous advancement and improvement of scientific and technological research, the number of scientific and technological papers, fund projects and patent works has shown an explosive growth trend[1]. These multi-type, multi-form and universally connected data constitute a heterogeneous information network（HIN）[2] with complex structure, large scale and interconnection. Shortly, HIN is a directed graph with multi-type nodes and different connection relationships. At the same time, network embedding[3][4], an effective method to represent large-scale networks, can well preserve the proximity of rich semantic information in original networks.

In order to combine the advantages of HIN with network embedding, the embedding representation learning of HIN has attracted extensive attention in scholar network[5][6]. It is common practice to use meta path, a sequence of relationships representing connections between various types of nodes, to capture the topological and associated semantic information of nodes in HINs. The embedding results of HIN have also been proved to be useful as feature input and widely used in various graph analysis tasks, especially clustering[7][8], classification[9][10], link prediction[11][12] and recommendation[13][14]. Although many scholars have proposed to learn meta-path-based node embedding representation methods in HINs, there still exist some unsolved challenges so far.

First, existing HIN embedding methods usually assume that the meta path weights of all nodes are the same, which cannot capture the individualized feature of nodes connected by path instances which represent different meta paths, leading to the insufficiency of degree of proximity preservation of the network topology. Therefore, it is necessary to effectively capture the personalized semantic feature of meta paths to help learn better node representations in HINs. In additional, while measuring node similarity, most of the previous methods only count the number of path instances, which lead to the ignorance of semantic differences between path instances. Taking a specific meta path in scholar-academic heterogeneous network as an example, meth path "author-paper-author" means that different scholars have published the same paper in the relationship of co-authors, that is to say, one author can connect to other authors through this meta path instance. Whereas meta path "author-paper-conference-paper-author" indicates there are two authors who have published their own paper at the same conference. An obvious fact is that to learn better node representations, we should distinguish these different path instances and focus on the most relevant ones while ignoring the noisy ones. Unfortunately, the ability of previous methods to capture such personalized preferences is insufficient.

At the same time, with data becoming a factor of production and playing an increasingly important role,

some traditional methods, such as graph neural networks and support vector machines, are based on all data for centralized training. However, due to the competition and monopoly between industries, the business of different enterprises is limited by certain commercial factors[15] and it is difficult to fully share data[16], which makes the exchange of data and information between enterprises difficult, and there is a serious data island phenomenon. To solve the above problems, in 2016, Google firstly proposed theory related to federated learning[17]. As an emerging paradigm of machine learning, it provides a novel solution for user data sharing, enabling users to obtain a more optimized model without the need for local original data, so that "data does not move and model moves". On the premise of ensuring data security and user privacy, it can not only break data island, but also fully tap the potential value of data.

To tackle above challenges, we proposed an attention-based representation model for academic heterogeneous information network embedding that can simultaneously preserve the rich node topology information and meta-path-based semantic information in HINs. We combined federated learning with embedded representation learning of HINs composed of scientific research teams and focus on optimization method of representation learning model based on federated training mechanism. The main contributions of this paper are summarized as follows:

(1) We constructed an academic heterogeneous information network to address the issue that large variety of scientific and technological data with complex structure and hidden semantics cannot be accurately represented by traditional homogeneous information network.

(2) We proposed an academic heterogeneous information network embedding representation learning method based on federated learning (FedAHE), which utilizes node attention and meta path attention mechanism to preserve both the topological and meta-path-based semantic information of nodes in HINs.

(3) We combined federated learning with the embedded representation learning of HINs to break the barriers of data islands, and design a federal training mechanism based on dynamic weighted aggregation of parameters (FedDWA) to optimize the representation learning method of HINs.

## 2 Related Work

In this section, we discuss some related research about heterogeneous information network embedding and federated learning.

***Heterogeneous Information Network Embedding*** As an effective paradigm for modeling complex relationships and objects[18][19][20], heterogeneous information network has attracted extensive research in emerging research directions.

Some scholars try to combine the representation method of homogeneous information network with the characteristics of heterogeneous information network[21][22][23][24][25]. For example, Grover[21] et al. designed node2vec to perform a biased walk by constraining the depth and breadth of the walk, and finally used the Word2vec[22] method for network representation learning. The representation vector of more structural information, although the accuracy of this method in the node classification experiment has been improved, but this method is only suitable for the representation learning of homogeneous networks. Based on game theory, GraphGan[23] et al. used the generator to fit the continuous distribution of nodes on all other nodes in the network, and generated the nodes most likely to be connected to the nodes; the discriminator distinguishes the generated node pairs from the real node pairs, through the minimax game finally fits the true nodes distribution and obtains the vector representation of network nodes[26][27]. This method can also be used for the representation of heterogeneous networks while ignoring the extra properties of nodes in heterogeneous networks, but the performance is mediocre. Meilian[28] et al. proposed a vector element path model, which embeds different types of nodes into different spaces to deepen the differences between these multi-type nodes, and further proposed an improved Metapath2vec++ model. Liu[29] et al. combined various types of objects with their relationships under a unified framework to calculate the heterogeneous reachability probability using a meta-path-based outlier identification technique in the representation learning of HINs.

***Federated Learning*** As well known, the scale of heterogeneous information networks is getting larger and larger as a result of the continuous advancement and improvement of data collecting technique[30][31]. Traditional information network representation methods usually construct high-dimensional sparse matrices, which require high computing and storage costs, and data requires centralized training and lacks flexibility[32][33]. At the same time, due to factors such as competition among industries, the representation learning method of traditional heterogeneous information networks will not be able to break the data island[34][35], and it is difficult to achieve full data sharing[36][37]. Based on this, federated learning emerges as the times require, which allows multiple data owners. Collaborative model building without compromising data privacy, and related research in the privacy protection mechanism of federated learning is rapidly expanding. Most studies use secure multi-party computation[38], homomorphic encryption[39], and differential privacy[40] to improve security at the same time. The aggregation mechanisms of parameters on federated learning server has also received extensive attention. The two primary categories of these works are exponential moving average based algorithm and federated average based algorithm. Taking an average aggregated weights of all trained weights in worker nodes participated in federated is what the original federated averaging algorithm[41] aims at. This means faster nodes must have to wait for the slower ones. In response to this problem, FedProx[42] dynamically adjusts local training rounds on various performance working nodes by a parameter server which is responsible to restrict the weight provided to itself by working nodes during local multi-round training and decrease the difference between slow and fast nodes. However, FedProx's method of directly weighting the corresponding positions of the working node weights reduces the convergence[43]. Moreover, the exponential moving average aggregation update strategy[44] has high efficiency in the case of high communication delay and heterogeneity since it can be weighted and summed with the weight reserved by the parameter server with a fixed weight, but different worker nodes send different versions of weights with a fixed weighting ratio will also reduce the convergence performance.

# 3 Preliminaries

In this section, we introduce several definitions related to our work and define our problem.

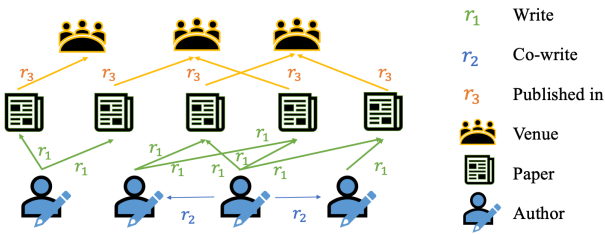

**Figure 1** A Heterogeneous Information Network in Sci-Technological Field

***DEFINITION 1 (Heterogeneous Information Network).*** Heterogeneous information network is a directed graph $G = (V, E, \varphi, \psi)$ where $V$ and $E$ respectively denotes the collections of the multi-type entity nodes and relational links, $\varphi: V \to A$ and $\psi: E \to R$ respectively denotes a map function of node type and link type which satisfy the condition that $\varphi(v) \in A$ and $\Psi(e) \in R$ for each object $v \in V$ and each link $e \in E$

***EXAMPLE 1.*** As the heterogeneous information network in sci-technological field shown in Figure 1, there are three kinds of nodes (eg: Venue, Paper, Author) and three kinds of edges which represent relationships between these nodes (eg: Write, Co-Write, Published in). These multi-type, multi-form and universally connected nodes constitute this heterogeneous information network with complex structure, large scale and interconnection.

***DEFINITION 2 (Network Schema and Meta Path).*** Network schema $T_G = \{A, R\}$ of a given HIN $G = (V, E, \varphi, \psi)$ is a directed graph defined on object and relation types, where $A$ and R respectively denotes the collections of object types and relation types. Thus network schema can be regarded as a meta description of HINs. Another important concept in heterogeneous information network analysis is meta paths. A meta-path is a path defined on a network schema that links two types of objects, formally denoted as $\pi = A_1 \longrightarrow A_2 \longrightarrow \cdots \longrightarrow A_m$ .

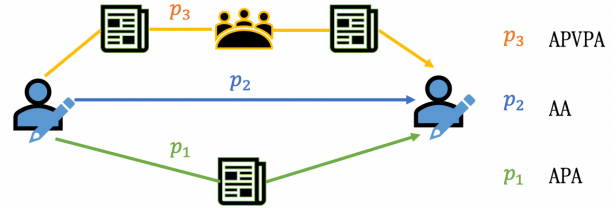

**Figure 2** Path Instances Between Two Author-nodes

***EXAMPLE 2.*** Figure 2 describes three possible path instances with different lengths that connect two authors. There we denote a path instance by using the first letter of each node through the meta path. For example, Author-Paper-Author is denoted as APA.

***DEFINITION 3 (Federated Learning).*** Federated learning, a machine learning technique, enables a set of organizations or divisions within the same organization to train and enhance a shared, global machine learning model iteratively and cooperatively with local dataset. Significantly different from traditional centralized machine learning methods, this method no longer exchanges and shares all local data with a central server.

***PROBLEM DEFINITION (Information Network Embedding).*** Given a HIN $G = (V, E, \varphi, \psi)$, a map function of node types $\varphi: V \longrightarrow A$, a map function of edge types $\psi: E \to R$ .What the HIN embedding representation aims to optimize is a map function $f: V \longrightarrow R_d$ which maps node $v \in V$ into a d-dimensional vector, where $R_d$ denotes the d-dimensional space.

# 4 The Proposed Method

In this section, we show the details of our proposed embedding representation learning framework of HINs based on federate learning. Figure 3 shows the architecture of our framework. In addition, the architecture of local HIN embedding model is shown in Figure 4.

Our model is designed as a central parameter server and a set of local clients. Different clients maintain their own HIN without sharing entity and relationships each other for training and updating respective HIN's embeddings. The central server is responsible for dynamical aggregating the overall weight parameters from different clients.

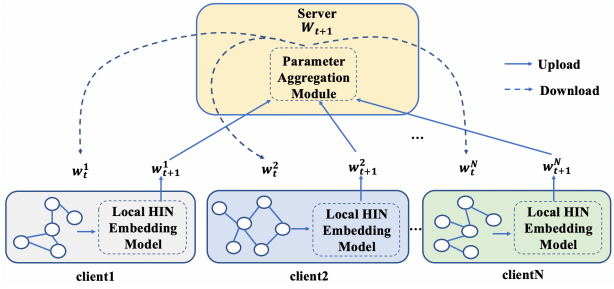

**Figure 3** Federated Academic HIN Embedding Model Architecture (FedAHE)

## 4.1 Client Update

Different clients should locally train and update their own HIN's embeddings. We design the local HIN embedding model with two layers, the node lever attention layer commits to learn structural feature embeddings of nodes in HINs by combing the structural feature representation of nodes itself and its neighbor nodes, and the meta-path level attention layer aims at learning semantic feature representation of nodes in HINs by capturing the rich semantic features carried by different meta paths and distinguishing the importance and preference of these meta paths. Thereby, a comprehensive embedding representation of nodes in HINs can contain both structural and semantic features. The local HIN Embedding model shown in Figure 4.

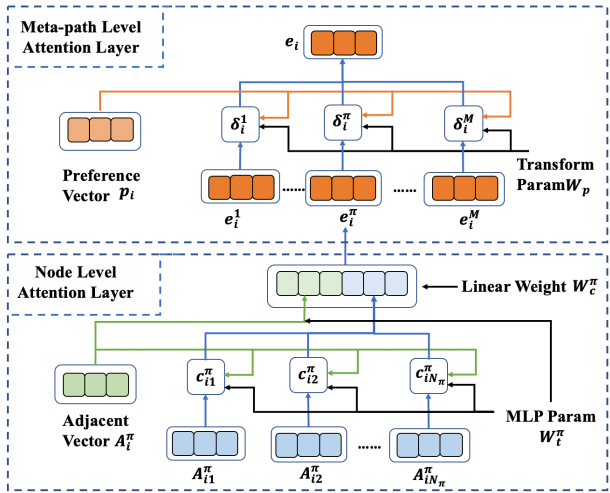

**Figure 4** Local HIN Embedding Model

### 4.1.1 Node Level Attention Layer

A basic idea to learn the structural feature representation of nodes in HINs is to combine the structural feature representation of nodes itself and its neighbor nodes. Moreover, those neighbors with similar structure to the node will be assigned a large node level attention weight.

Considering that nodes in different semantic spaces should have different structural features, so that the importance of different meta paths can be better distinguished, we learn the embedding representation of node structural features based on adjacent vector $A_i^\pi$. There is no doubt that the adjacent vector based on meta path could be high-dimensional, thus a multi-layer perceptron (MLP), parameterized by $W_t^\pi$, is applied to transfer the high-dimensional sparse adjacent vector into a d-dimensional space.

For the neighbors $V_j$ of node $V_i$, $V_j$ will be assigned a larger node-level attention weight if it has more similar structure to the node $V_i$, here cosine similarity is applied to measure the similarity of structural features between $V_i$ and $V_j$ and the node-level attention coefficient $c_{ij}^\pi$ can be computed as:

$$s_{ij}^\pi = \frac{(W_t^\pi A_i^\pi)^T \cdot W_t^\pi A_j^\pi}{\|W_t^\pi A_i^\pi\| \cdot \|W_t^\pi A_j^\pi\|} \qquad (1)$$

$$c_{ij}^\pi = \frac{\exp(s_{ij}^\pi)}{\sum_{k \in N_i^\pi}(s_{ik}^\pi)} \qquad (2)$$

where $W_t^\pi$ is the parameter of MLP which denotes the transformation matrix for meta path $\pi$, $s_{ij}^\pi$ denotes the meta-path-based cosine similarity of structural feature between $V_i$ and $V_j$, $c_{ij}^\pi$ is the node-level attention coefficients of node $V_j$ and $N_i^\pi$ denotes the collections of neighbors connecting to node $V_i$ by meta path $\pi$.

To learn the aggregated structural feature embedding representation $e_{N(i)}^\pi$ of nodes $V_i$ by uniformly sampling nodes in the set of its neighbor nodes:

$$e_{N(i)}^\pi = \sigma\left(\sum_{j \in N_i^\pi} \alpha_{ij}^\pi W_t^\pi A_j^\pi\right) \qquad (3)$$

where $e_{N(i)}^\pi$ denotes the aggregated embedding of node $V_i$, $N_i^\pi$ is the neighborhood nodes connected to node $V_i$ by meta path $\pi$ and $\sigma()$ is the activation function.

Finally, by combing the structural feature representation of nodes itself and its neighbor nodes connected by path instances, the meta path based embedding $e_i^\pi$ of $V_i$ can be computed as follow:

$$e_i^\pi = W_c^\pi[e_{N(i)}^\pi \cdot W_t^\pi A_i^\pi] \qquad (4)$$

where $e_{N(i)}^\pi$ is the aggregated structural feature embedding of node $V_i$, $W_t^\pi A_i^\pi$ is the structural feature representation of nodes $V_i$, $[\cdot]$ indicates vector concatenation operation, and $W_c^\pi$ denotes the dimensional transformation weight matrix.

### 4.1.2 Meta-path Level Attention Layer

Based on the effectively learning in node level attention layer, we get the structural feature embedding representation $\{E^1, E^2, ..., E^m\} \in R^{N \times d}$ which persevere the structural similarity of these nodes. In order to better capture the rich semantic features carried by different meta paths and distinguish the importance and preference of these meta paths to each node, we design the meta-path level attention mechanism based on a meta-path preference vector $p_i \in R^{1 \times k}$ .Thereby, a comprehensive embedding representation of nodes in HINs can contain both structural and semantic features.

Similarly, a more similarity between the structural feature embedding $e_i^\pi$ based on meta path and the preference vector $p_i$ will lead to larger attention coefficients. To measure the cosine similarity between $e_i^\pi$ and $p_i$ ,we first apply a line transformation parameterized by $W_p$ to transform the structural feature embedding $e_i^\pi$ into k-dimension space and denoted the transformed embedding as $e_i^{\pi'}$ ,thus the meta path attention coefficients of node $V_i$ can be denoted as:

$$\delta_i^{\pi'} = \frac{p_i^T \cdot e_i^{\pi'}}{\| p_i \| \cdot \| e_i^{\pi'} \|} \qquad (5)$$

$$\delta_i^\pi = \frac{\exp (\delta_i^{\pi'})}{\sum_{m=1}^M \exp (\delta_i^{\pi^m})} \qquad (6)$$

where $p_i$ is the preference vector, $\delta_i^{\pi'}$ is the cosine similarity between the transformed structural feature embedding $e_i^{\pi^i}$ based on meta path, $\| \cdot \|$ denotes vector L2 normalization and $\delta_i^\pi$ is the meta path attention coefficients of node $V_i$ .

Finally, given a HIN in which $M$ meta paths exist, the ultimate comprehensive embedding representation $e_i$ of node $V_i$ which well preserves both rich semantic information and structural feature in HIN can be denoted as:

$$e_i = \sum_{\pi=1}^M \delta_i^\pi e_i^\pi \qquad (7)$$

The objective of client updates is to minimize the cross-entropy loss between the predictions and ground-truth by back propagation and stochastic gradient descent. The loss function is denoted as follow:

$$L = -\sum_i^{|V|} \sum_{l=1}^L Y_l^l \ln (S(E_i)^l) \qquad (8)$$

where $S(E_i)^l \epsilon \{0,1\}$ and $Y_l^l \epsilon \{0,1\}$ respectively denotes the predicted label and ground truth of node $V_i$ which belongs to label $l$ .

## 4.2 Server Aggregation

In our model, different clients train and update their own respective HIN's embeddings by the local HIN embedding model. Meanwhile, the server is responsible of dynamical aggregating the overall weight parameters from different clients and sends aggregated parameters to them.

Considering that the learning efficiency and convergence speed of the local model in federated learning have certain differences due to factors such as the differences of device performance. As a result of the fast working nodes' frequent parameter submissions, the aggregated model parameters in this situation diverge from the convergence direction of other nodes in the model, and the slow nodes' sluggish parameter submissions impede the convergence of the parameter server model.

To solve above problems, we maintain two records on the federated parameter server. A record of the latest weight parameters of all clients participating in the federation is saved on the sever to guarantee that while

---

**Algorithm 1: FedAHE Framework**

**Input**: HIN $G = (V, E, \varphi, \psi)$; Embedding dimension $d$ and preference vector dimension $k$; Meta path set $\{\pi_1, \pi_2, \dots, \pi_M\}$; Number of federated clients $C$; Local epoch number $e$ and local batch size $B$.
**Output**: Node embedding vector $E$.

**Server Aggregation:**
    Initial and save version and weight record.
    **for** round r=0,1,$\cdots$ **do**
        Aggregate and update uploaded parameters $w_{latest\_normal}$ by Eq. (9) and Eq. (10).
        **If** version gap < threshold **then**
            Only send aggregated parameters to client with id = $id$.
        **else**
            Send aggregated parameters to all clients.
        **end if**
    **end for**

**Client Update:**
    Randomly initialize transformation matrix $W_t$, $W_c$, $W_p$ and preference vector $p_i$ for each node $V_i$ .
**for** each local epoch from 1 to $e$ **do**
    Random shuffle nodes into batches of size $B$.
    **for** each batch $b$ **do**
        **for** $\pi \epsilon \{\pi_1, \pi_2, \dots, \pi_M\}$ **do**
            Compute node level embedding $E$ by Eq. (1), Eq. (2), Eq. (3) and Eq. (4).
        **end for**
        **for** $\pi \epsilon \{\pi_1, \pi_2, \dots, \pi_M\}$ **do**
            Compute meta-path level embedding $E$ by Eq. (5), Eq.(6) and Eq.(7).
        **end for**
        Compute loss $L$ for $b$ by Eq.(8).
        Upload parameters to server.
        Download aggregated parameters from server and update model parameters.
    **end for**
**end for**
Return node embedding $E$.

updating clients, all out-of-date weights of federated clients can be totally eliminated. We denote the weight records at $t$ on the server as $S_w = \langle w_1^t, w_2^t, \ldots, w_n^t \rangle$. Moreover, a record of the latest parameter version number all worker nodes is also preserved, and denoted the version records at $t$ on the server as $S_v = \langle v_1^t, v_2^t, \ldots, v_n^t \rangle$.

Thanks to these two records, during aggregation, according to the version discrepancy between the most recent version offered on various client nodes and the current latest version recorded on the parameter server, we can dynamically increase the weight of nodes with a small version gap and decrease the weight of the nodes with a large version gap. The overall producer of our framework shows in Algorithm 1, for more details, we describe the federated parameters dynamic-weight aggregation (FedDWA) in Figure 5.

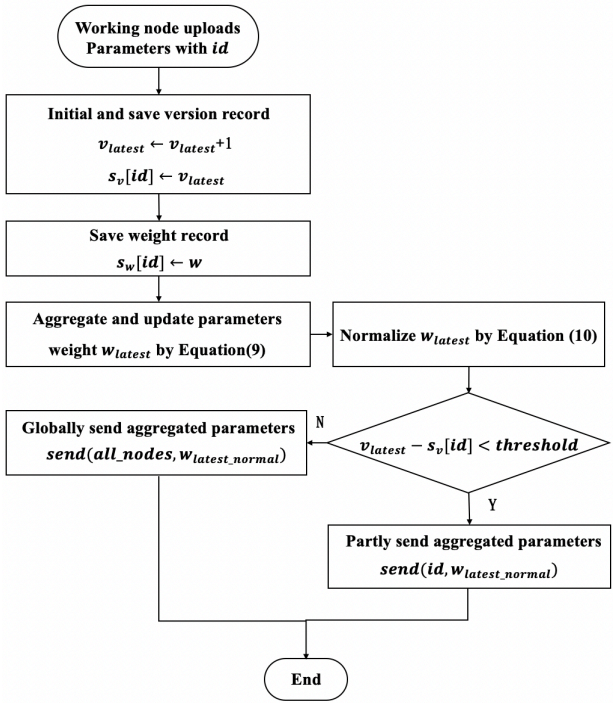

**Figure 5** The Algorithm Flowchart of Federated Parameters Dynamic-weight Aggregation (FedDWA)

Suppose there are $n$ working nodes each identified by a unique $id$ number in federated joint training, the version number is initialized to 1 and increments by 1 after each update occurs. When each time uploading model parameters, each node sends its own node id, current weight $w$ and version number $v_{latest}$ to the parameter server and only the record $S_w[id]$ of node with $id$ is updated. The aggregated parameter weights $w_{latest}$ of working node with $id$ are updated as follows:

$$w_{latest} = \sum_i^n ((v_{latest} - S_v[id] + 1)^{-\alpha}) \cdot S_w[id] \quad (9)$$

where $v_{latest}$ is the latest version provided by working node with $id$, and $S_v[id]$ is the corresponding version record saved on the parameter server.

Moreover, the normalization aggregated parameter weights $w_{latest\_normal}$ is denoted as:

$$w_{latest\_normal} = \frac{w_{latest}}{\sum_i^n ((v_{latest} - S_v[id] + 1)^{-\alpha})} \quad (10)$$

Each time the aggregation is completed, it is checked whether the version of all nodes is behind the latest version, and whether it exceeds the threshold limit of the version gap. When the version gap limit is exceeded, the parameter server will send the latest weight to all nodes, otherwise the updated aggregation parameters will be only sent to this client with id= $id$.

# 5 Experiments

We conduct experiments on three real-world HINs (DBLP[45], AMiner[46] and ACM[47]) oriented scientific research to prove the correctness of our proposed framework. These three datasets, typical representative of heterogeneous information network, contain information about papers, paper citation relationships, authors and author collaboration. The datasets' specifics are discussed below and Table I includes a summary of their statistics.

**Table I** Dataset Statistic

|  | Node | Edge | Label | Meta Path |
|------|------|------|-------|-----------|
| DBLP | 10650 | 39888 | 4 | APA, APPA, APTPA, APVPA |
| Aminer | 12480 | 30736 | 2 | APA, APPA |
| ACM | 8916 | 37646 | 3 | APA, APPA |

## 5.1 Baseline

To demonstrate the capability of learning heterogeneous information in HINs, we compare to Node2Vec and LINE which loss network heterogeneity since regarding all nodes and relationships as the same type. Also, we compare to Metapath2vec, HIN2vec and GAT for providing evidence about the superiority in heterogeneity exploring of our proposed attention mechanism.

**Node2vec/LINE[48][49]** Node2vec respectively applies skip-gram and truncated random walks model for generating node sequences and learning node embeddings. LINE preserves both first order and second order proximity to learn node embeddings by using the connection relationship of nodes to determine the empirical distribution.

**Metapath2Vec[50]** Its core idea is to perform a random walk with consideration of meta path for node sequence gain, then utilize skip-gram to solve the vectorized solution.

**HIN2vec[51]** It also takes meta path into account when learning the embedding of nodes based in HINs while ignoring the weight of different meta paths.

**GAT**[52] GAT, a graph attention based network embedding method, uses masked self-attention layers to obtain the neighborhood features of each node by stacking network layers, and assigns different weights to different nodes in the neighborhood.

For experimental implementation, we apply Adam optimizer with initial learning rate as 0.001 for local HIN Embedding model optimization. In addition, we respectively set the local batch size and epoch to {64,128,256} and {1,3,5}. Moreover, the embedding vector dimension is 128 and the number of clients chosen is 3.

## 5.2 Node Classification

We choose Macro-F1 and Micro-F1 score as the evaluation metric for node classification, which is widely applied to evaluate network embedding results. The results of node classification in datasets mentioned above are shown in Table II.

According to the first two rows of the table, the metrics of our proposed method is higher than that of all these homogeneous network embedding methods (Node2Vec and LINE), which indicates that our embedding method has a more efficient ability and better performance to correctly capture the rich heterogeneous information in HINs. For the next two rows representing methods (MetaPath2Vec and HIN2Vec) that only consider meta-paths, but ignore the preference weights of these different meta-paths, our proposed method almost performs best. Furthermore, our proposed methods also perform better than GAT which takes the weights of different meta path in account, demonstrating the efficiency, correctness and feasibility of proposed node level attention and meta-path attention mechanism.

visualize node embeddings learning result by t-SNE algorithm.

The results of network visualization obtained by applying the top-3 embedding methods on DBLP are shown in Figure 6. As the figure shows, by effectively learning meta-path-based semantics and topological similarity of nodes, our proposed approach offers denser clusters with more distinct category boundaries than others.

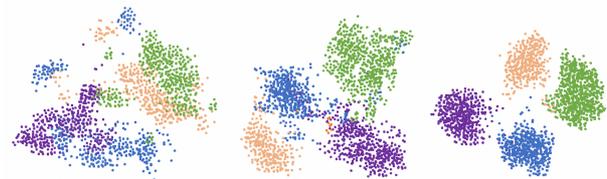

(a)MetaPath2Vec (b)HIN2Vec (c) FedAHE(ours)

**Figure 6** The Results of Network Visualization Obtained by Applying the Top-3 Embedding Methods on Dataset DBLP

## 5.4 Parameter Analysis

In this part, we commit to explore the impact of several key parameters on convergence rate such as the client computation of local model training (local epoch number $e$ and batch size $B$) and the communication rounds between clients and server.

Due to different federated clients may have different computation power which lead to different convergence rate, in our federated HIN embedding model, client computation is thus a significant component that should be taken into account. Client computation can usually be denoted by $e/B$, where $e$ is the local epoch number and $B$ is the batch size.

**Table II** Node Classification on Three Academic HIN Dataset

| Dataset | DBLP | | Aminer | | ACM | |
|---|---|---|---|---|---|---|
| Evaluation | MicroF1 | MacroF1 | MicroF1 | MacroF1 | MicroF1 | MacroF1 |
| Node2Vec | 0.836 | 0.837 | 0.841 | 0.826 | 0.874 | 0.865 |
| LINE | 0.841 | 0.843 | 0.899 | 0.881 | 0.913 | 0.896 |
| MetaPath2Vec | 0.783 | 0.776 | 0.923 | 0.878 | 0.921 | 0.887 |
| HIN2Vec | 0.850 | 0.852 | 0.927 | **0.902** | 0.927 | 0.904 |
| GAT | 0.843 | 0.840 | 0.921 | 0.899 | 0.867 | 0.786 |
| FedAHE(ours) | **0.867** | **0.859** | **0.932** | 0.901 | **0.931** | **0.913** |

## 5.3 Network Visualization

Generally, network visualization can provide us the possibility to intuitively understand node embedding results. To intuitively show that our proposed framework can more effectively learn both topological and semantic features of nodes, we

We experiment on DBLP with fixed client number $C$ as 3, enumerating local epoch $e$ with {1,3,5} and local batch size $B$ with {64, 128, 256} to explore how client computation effects rate of model convergence. Figure 7 and Figure 8 show our results.

According to the curves in Figure 7, we can find that smaller local epoch can improve model performance but slowing down the model convergent rate. Similar to this, Figure 8 demonstrates that larger local batch size makes better model performance but reducing model convergent rate. And it is worth mentioning that our model preforms best with local epoch $e = 1$ and local batch size $B = 256$.

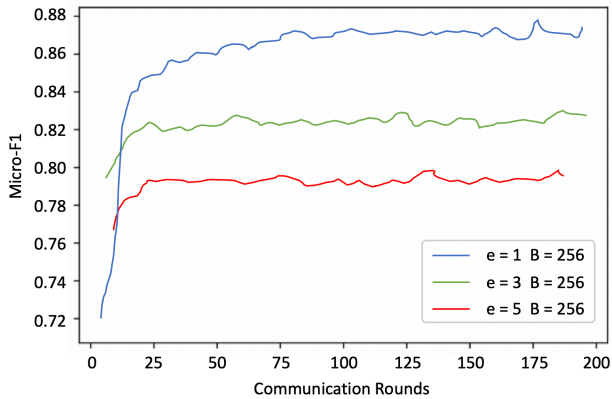

**Figure 7** Curves with fixed local batch size while enumerating local epoch on DBLP

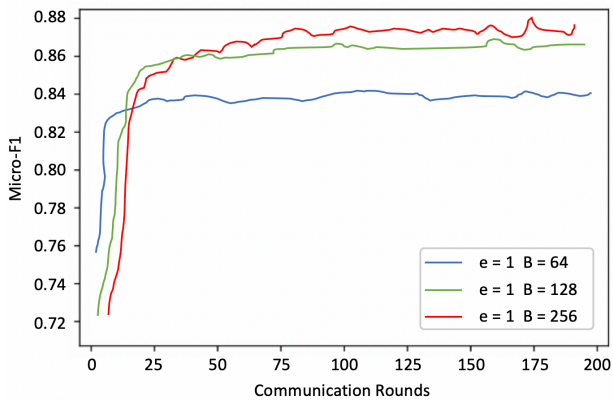

**Figure 8** Curves with fixed local epoch while enumerating local batch size on DBLP

### 5.5 Parameter Aggregation

We also compare our proposed FedDWA with FedAvg (Federated Averaging) and AFed (Asynchronous Federated Learning based on exponential moving average), two typical federated parameter aggregation algorithms.

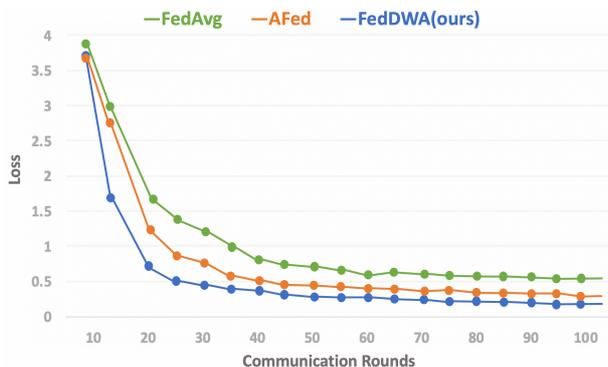

**Figure 9** The Rate of Gradient Descent with Different Parameter Aggregation Methods on DBLP

Figure 9 shows the rate of gradient descent with different parameter aggregation methods on DBLP. We can find that the gradient decreases faster on the same communication round, and the model also converges faster than the traditional methods, demonstrating that the proposed FedDWA works effectively by dynamic aggregating and updating parameter weights uploaded by federated participants.

## 6 Conclusions

We proposed a federated embedding representation learning method of academic heterogeneous information network (FedAHE). This method apply meta path attention and node attention mechanism, which can not only learn low-dimensional, dense vector representations of nodes in HINs, but also preserve the rich network topological information and meta-path-based semantic information of nodes. In additional, we combined federated learning with embedding learning of HINs, proposed a federal training mechanism based on dynamic weighted aggregation of parameters (FedDWA) to optimize the local HINs embedding learning. Through experiments, the proposed method has been proved to have much better correctness, feasibility and efficiency than the traditional methods.

## 7 Acknowledgements

This work was supported by the National Natural Science Foundation of China (No.62192784, No.62172056, No.61877006).